\documentclass[10pt,conference]{IEEEtran}
\usepackage{times,booktabs,multirow}
\usepackage{latexsym,subfigure}
\usepackage{amsmath,amsfonts} 
\usepackage{tikz,amsmath,amsthm,graphicx,subfig,makecell}
\usepackage{algorithmic,algorithm,varwidth,tabularx,cellspace,caption}
\usepackage{url}

\begin{document}

\title{\huge{The Role of User Profiles for Fake News Detection}}

\author{
\IEEEauthorblockN{
Kai Shu\IEEEauthorrefmark{1},
Xinyi Zhou\IEEEauthorrefmark{2}
Suhang Wang\IEEEauthorrefmark{3},
Reza Zafarani\IEEEauthorrefmark{2}, and
Huan Liu\IEEEauthorrefmark{1}
}
\IEEEauthorblockA{
\IEEEauthorrefmark{1}Arizona State University, \{kai.shu, huan.liu\}@asu.edu\\
\IEEEauthorrefmark{2}Penn State University, szw494@psu.edu\\
\IEEEauthorrefmark{3}Syracuse University, \{zhouxinyi, reza\}@data.syr.edu}}

\maketitle


\begin{abstract}

Consuming news from social media is becoming increasingly popular. Social media appeals to users due to its fast dissemination of information, low cost, and easy access. However, social media also enables the widespread of \textit{fake news}. Because of the detrimental societal effects of fake news, detecting fake news has attracted increasing attention. However, the detection performance only using news contents is generally not satisfactory as fake news is written to mimic true news. 
Thus, there is a need for an in-depth understanding on the relationship between \textit{user profiles} on social media and fake news. 
In this paper, we study the challenging problem of \textit{understanding} and \textit{exploiting} user profiles on social media for fake news detection. In an attempt to
understand connections between user profiles and fake news, first, we measure users' sharing behaviors on social media and group representative users who are more likely to share fake and real news; then, we perform a comparative analysis of explicit and implicit profile features between these user groups, which reveals their potential to help differentiate fake news from real news. To exploit user profile features, we demonstrate the usefulness of these  user profile features in a fake news classification task. We further validate the effectiveness of these features through feature importance analysis. The findings of this work lay the foundation for deeper exploration of user profile features of social media and enhance the capabilities for fake news detection. 
\end{abstract}
\maketitle

\section{Introduction}
Due to the increasing amount of time spent on social media,  people increasingly tend to seek out and receive their news through social media sites. In December 2016, the Pew Research Center announced that approximately 62\% of US adults get news from social media in 2016, while in 2012, only 49\% reported reading news on social media.\footnote{http://www.journalism.org/2016/05/26/news-use-across-social-media-platforms-2016/} This rapid rate of increase in user engagements with online news can mainly be attributed to the cheap, mobility, and fast dissemination of social media platforms.
However, despite these advantages, the quality of news on social media is considered lower than that of traditional news outlets. Every, large volumes of fake news, i.e., news stories with intentionally false information~\cite{shu2017fake}, are widely spread online. For example, a report estimated that over 1 million tweets were related to the fake news story ``Pizzagate''\footnote{\url{https://en.wikipedia.org/wiki/Pizzagate_conspiracy_theory}} by the end of 2016 presidential election.

Fake news has several significant negative societal effects. First, people may accept deliberate lies as truths~\cite{nyhan2010corrections}. Second, fake news may change the way people respond to legitimate news.  A study has shown that people's trust in mass media has dramatically degraded across different age groups and political parties~\cite{swift2016americans}. Finally, the prevalence of fake news has the potential to break the trustworthiness of the entire news ecosystem. Thus, it is critical to detect fake news on social media to mitigate these negative effects, and benefit the general public as well as the entire news ecosystem.

However, detecting fake news on social media presents unique challenges. First, fake news is intentionally written to mislead readers, which makes it nontrivial to detect simply based on content; Second, social media data is large-scale, multi-modal, mostly user-generated, sometimes anonymous and noisy. Recent research advancements aggregate uses profiles and engagements on news pieces to help infer articles that are incredible~\cite{castillo2011information}, leading to some promising early results. However, no principled study is conducted on characterizing the profiles of users who spread fake/real news on social media. In addition, there has been no research that provides a systematic understanding of (i) what are possible user profile features; (ii) whether these features are useful for fake news detection; and (iii) how discriminative these features are. To give a comprehensive understanding, we investigate the following three research questions:



\begin{itemize}
  \item \textbf{RQ1:} \emph{Which users are more likely to share fake news or real news?}
  \item \textbf{RQ2:} \emph{What are the characteristics/features of users that are more likely to share fake/real news, and do they have clear differences?}
  \item \textbf{RQ3:} \emph{Can we use user profile features to detect fake news and how?}
\end{itemize}
By investigating \textbf{RQ1}, we aim to identify users who are more likely to share fake or real news, which can be treated as representative user sets to characterize user profiles. By answering \textbf{RQ2}, we can further provide guidance on assessing whether the profiles of identified users are different or not, and to what extent and in what aspects they are different. In addition, by studying \textbf{RQ3}, we explore different ways to model user profile features, analyze the importance of each feature and show the feature robustness to various learning algorithms. By answering these research questions, we made the following contributions:

\begin{itemize}
  \item We study a novel problem of understanding the relationships between user profiles and fake news, which lays the foundation of exploiting them for fake news detection;
  \item We propose a principled way to characterize and understand user profile features. We perform a statistical comparative analysis of these profile features, including explicit and implicit features, between users who are more likely to share fake news and real news, and show their potentials to differentiate fake news; and 
  \item We demonstrate the usefulness of the user profile features to classify fake news on real-world datasets, whose performance consistently outperforms existing state-of-the-art features extracted from news content. We also show that the extracted user profile features are robust to different learning algorithms, with an average $F1$ above 0.90.
 We further validate the effectiveness of these features through feature importance analysis, and found that implicit features, e.g., political bias, perform better than explicit features.
  \end{itemize}

\section{Assessing Users' Sharing Behaviors}\label{sec:assessing}
We investigate \textbf{RQ1} by measuring the sharing behaviors of users on social media on fake and real news.  We aim to identify users who are more likely to share fake or real news, which can be further used to characterize discriminative features for fake news detection. 

\subsection{Datasets}
\begin{table}[t]
\centering \caption{The statistics of FakeNewsNet dataset}
\begin{tabular}{l|cc}
\toprule
 Platform &Politifact & Gossipcop  \\
\midrule
\makecell{\# Users (without \\filtering botss)} &  159,699 & 209,930\\
\midrule
\# Sharing & 271,462 & 812,194\\
\midrule
\# True news& 361 & 4,513\\
\midrule
\# Fake news & 361 & 4,513\\
\bottomrule
\end{tabular} \label{tab:data}
\end{table}

We utilize one fake news detection benchmark data repository named  FakeNewsNet~\cite{shu2018fakenewsnet}. The datasets are collected from two   fact-checking platforms: \textit{Politifact}\footnote{https://www.politifact.com/} and \textit{Gossipcop}\footnote{https://www.gossipcop.com/}, both containing news content with labels annotated by professional fact-checkers, and social context information. News contents include meta attributes (e.g., body text), and social context includes the related user social engagements of news items (e.g., user posting/sharing news) on Twitter. The detailed statistics of the datasets are shown in Table~\ref{tab:data}.

\subsection{Filtering Bot Users}
Social bots have been reported to play an important role to spread fake news on social media~\cite{shu2017fake,davis2016botornot,ferrara2016rise}. The profiles of bots are usually manipulated to hide their identities~\cite{ferrara2016rise},  and may introduce noise on understanding the characteristics of user profile features. To alleviate the effects of social bots, we apply one of the state-of-the-art bot detection tool \textit{Botometer}\footnote{https://botometer.iuni.iu.edu}~\cite{davis2016botornot} to filter out bot accounts. Botometer takes a Twitter username as an input and utilizes various features extracted from meta-data obtained from Twitter API and outputs a probability in $[0,1]$, indicating how likely the user is a social bot. Following the common setting,  we filter out those users who have a score greater than 0.5. We keep the remaining users and treat them as authentic human users. 
Finally, we have filtered out \textbf{14.2\%} ($9,231$ out of $64,901$) and \textbf{13.7\%} ($11,257$ out of $82,163$) users for fake and real news on Politifact, and \textbf{21\%} ($26,879$ out of $127,446$)  and \textbf{18.9\%} ($11,854$ out of $62,516$) users for fake and real news on Gossipcop. We can see that a bigger ratio of bot users exist among those who spread fake news than those who spread real news.

\subsection{Identifying User Groups}

We identify different subsets of users based on their sharing behaviors on fake and real news. By finding these groups, we want to build representative user sets that are more likely to share fake/real news from which we can further compare the degree of the differences of their profiles to find useful profile features.  Towards answering \textbf{RQ1}, we propose two measures to assess user-news sharing behavior as following:

\subsubsection{Absolute Measure}\label{absolute}

To compare the sharing behavior of user $u_i$ with other users, we compute the absolute number of fake (real) news items that user $u_i$ has shared, denoted as $n^{(f)}_{i}$ ($n^{(r)}_{i}$). Intuitively, if users share more fake news compared with other users, they--as a population--tend to share fake news, and vice versa. Thus, we select those users who share the most absolute number of fake and real news, denoted as $\mathcal{U}^{(r)}_a\subset\mathcal{U}$ and $\mathcal{U}^{(f)}_a\subset\mathcal{U}$, where $\mathcal{U}$ is the set of all users. We compute $\mathcal{U}^{(r)}_a=\mbox{Top}K(n^{(r)}_{i})$, indicating the top-$K$ users that share the most real news pieces; and $\mathcal{U}^{(f)}_a=\mbox{Top}K(n^{(f)}_{i})$, indicating the top-$K$ users that share the most fake news pieces.

\subsubsection{Relative Measure}\label{relative}

Even some users may not share many fake news items relative to all users, they may still share more fake news in history. Thus, we propose a metric named Fake news Ratio (FR). $FR(i)$ denotes the FR score of user $u_i$ as $FR(i) = \frac{n^{(f)}_{i}}{n^{(r)}_{i}+n^{(f)}_{i}}$,
which is equivalent to the total number of fake news user $u_i$ has shared (i.e., $n^{(f)}_{i}$), divided by the total number of all news items he/she has shared (i.e., $n^{(r)}_{i}+n^{(f)}_{i}$). The larger the value, the higher the percentage of fake news items that are being shared by $u_i$. 

\subsubsection{User Groups} Based on the two measures, we introduce a principled way to identify representative user groups $\mathcal{U}^{(f)}$ and $\mathcal{U}^{(r)}$.
First, we divide all users into three subsets: (1) ``Only Fake'': users who only spread fake news; (ii) ``Only Real'': users who only spread real news; and (iii) ``Fake and Real'': users who spread both fake and real news, as shown in Table~\ref{tab:user_spreading_sta}.  Second, we empirically select top 10,000 users from ``Only Fake'' and ``Only Real'' ranked by the number of fake news or real news they share; and then we further select users with lower FR scores ($FR\in [0,t]$) and add to $\mathcal{U}^{(r)}$ with a threshold $t$; and select users with higher FR scores ($FR\in [1-t,1]$) and add them to $\mathcal{U}^{(f)}$. By changing the threshold of $t$, we find out we can obtain consistent results when $t<0.4$, and when $t\geq0.4$ more noisy tend to be included, and the comparison analysis may not be accurate. Thus we set $t=0.2$ to reduce the noise for the feature analysis. The selected users are equally sampled for both $\mathcal{U}^{(f)}$ and $\mathcal{U}^{(r)}$  (see Table~\ref{tab:user_spreading_sta}).
\begin{table}[tp!]
\centering
\caption{Users are categorized into three groups: ``Only Fake'', ``Only Real'' and ``Fake and Real'', according to whether they share only fake news, only real news, or both fake and real news.}
\begin{tabular}{l|cc}
\toprule
  & PolitiFact & GossipCop\\
 \midrule
 \# Only Fake & 57,926 & 112,697  \\
\midrule
 \# Only Real & 75,188 & 47,767\\
\midrule
 \# Fake and Real & 6,975 & 14,749 \\
\midrule
 \# Users & 140,089 &  175,213\\
 \midrule
 \# \textbf{Selected Users} & 10,684 & 11,785 \\
\bottomrule
\end{tabular} \label{tab:user_spreading_sta}
\end{table}

\section{Understanding User Profiles}\label{sec:charac}
Users in $\mathcal{U}^{(f)}$ are more likely to share fake news, and those in $\mathcal{U}^{(r)}$ are more likely to share real news. However, it is unknown to what extent and in what ways these users are different . Therefore, we explore \textbf{RQ2} to understand if there are clear differences among users in $\mathcal{U}^{(f)}$ and $\mathcal{U}^{(r)}$. In order to analyze the users from $\mathcal{U}^{(f)}$ and $\mathcal{U}^{(r)}$, we extract features for each set and compare them. 

We collect and analyze user profile features from different aspects, i.e., \textit{implicit} and \textit{explicit}.  Implicit features are not directly available but are inferred from user meta information or online behaviors, such as historical tweets. Explicit features are obtained directly from meta-data returned by querying social media site APIs. Our selected feature sets are by no means the comprehensive list of all possible features. However, we focus on those implicit features that are widely used in the literature for better understanding user characteristics, and explicit features that can be easily accessed and are available for almost all public users. We assume that: if a feature $f$ reveals clear differences between $\mathcal{U}^{(r)}$ and $\mathcal{U}^{(f)}$, then $f$ has potential usefulness for detecting fake news; otherwise, $f$ may not be useful for fake news detection. We will further test and verify this assumption in Section~\ref{sec:eval}. 

\subsection{Implicit Profile Features}
We explore several implicit profile features, which are not directly provided through user meta-data from Twitter API, but are widely used to describe and understand user demographics~\cite{schwartz2013personality}. Note that we adopt widely-used tools to predict these implicit features in an \textit{unsupervised} way. Our goal is to perform a fair comparison analysis on the predicted features, and the prediction accuracy of these features are not guaranteed and not the focus of this paper. 


\textbf{Age:} Studies have shown that age has impacts on people's psychology and cognition. For example, as age gradually changes, people typically become less open to experiences,  but more agreeable and conscientious~\cite{mccrae1999age}. Using this feature, we aim to answer whether users in different age groups have different abilities to differentiate fake news.

\begin{figure}[tp!]
\centering
\subfigure[Age on PolitiFact]{
  {\includegraphics[scale=0.25]{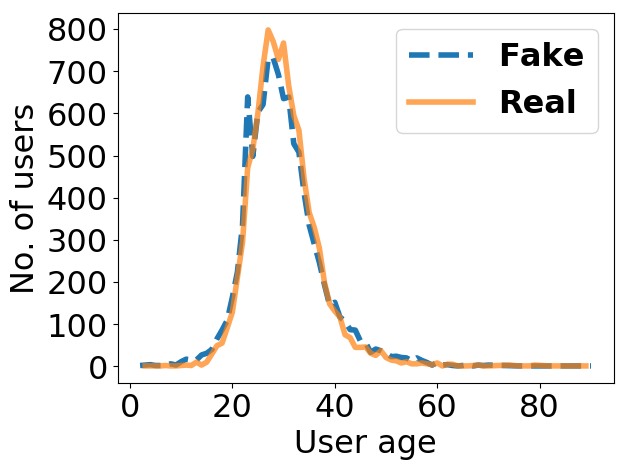}}
}
\subfigure[Age on GossipCop]{
  {\includegraphics[scale=0.25]{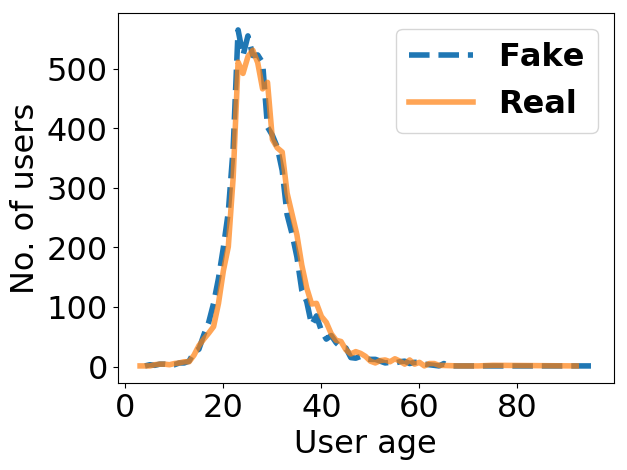}}
}
\caption{\textbf{Age Comparison}. We rank the ages from low to high and plot the number for users. The x-axis represents the predicted ages, and y-axis indicates the number of users.}\label{fig:age_gender}\vspace{-0.5cm}
\end{figure}

We infer the age using an state-of-the-art approach~\cite{sap2014developing}, which uses a linear regression model with the collected predictive lexica (with words and weights). We use user recent tweets as our corpus and extract relevant words in the lexica. The results are shown in Figure~\ref{fig:age_gender}.
We also perform the statistical $t$-test to verify whether predicted ages of users in $\mathcal{U}^{(f)}$ and $\mathcal{U}^{(r)}$ are significantly different or not, and the $p$-value$<$0.05, which indicates the results are statistically significant. From Figure~\ref{fig:age_gender}, we observe that  there are larger number of users who are predicted with older ages (e.g., after the peak point) spreading real news than those spreading fake news; and there are larger number of users who are predicted with younger ages (e.g., before the peak point) spreading fake news than those spreading real news, in both datasets.



\textbf{Personality:} Personality refers to the traits and characteristics that makes an individual different from others. We draw on the popular Five Factor Model (or ``Big Five''), which classifies personality traits into five dimensions: \textbf{E}xtraversion (e.g., outgoing, talkative, active), \textbf{A}greeableness (e.g., trusting, kind, generous), \textbf{C}onscientiousness (e.g., self-controlled, responsible, thorough), \textbf{N}euroticism (e.g., anxious, depressive, touchy), and \textbf{O}penness (e.g., intellectual, artistic, insightful). We try to answer the following question: do personality differences clearly exist between $\mathcal{U}^{(f)}$ and $\mathcal{U}^{(r)}$?
\begin{figure}[!bp]
\centering
\subfigure[Personality Distribution on PolitiFact]{
  {\includegraphics[scale=0.13]{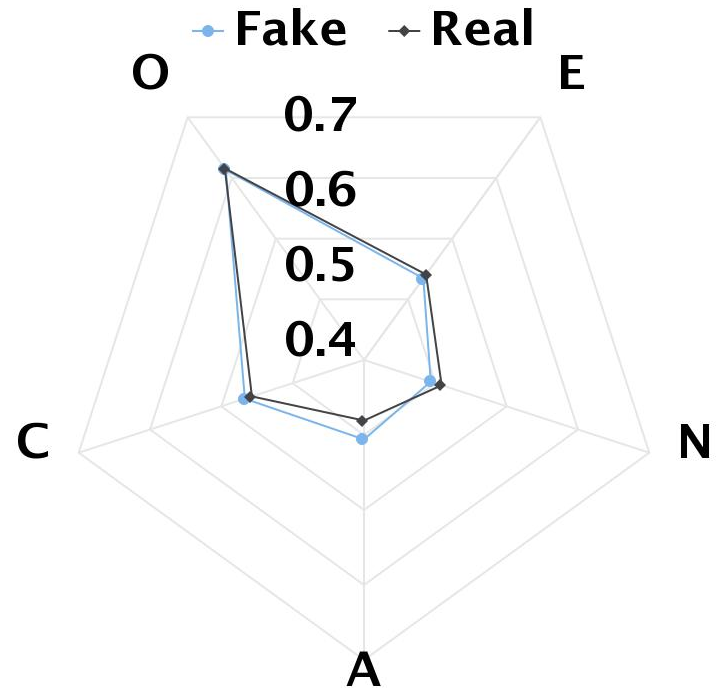}}
}
\hspace{-0.1cm}
\subfigure[Personality Distribution on GossipCop]{
  {\includegraphics[scale=0.13]{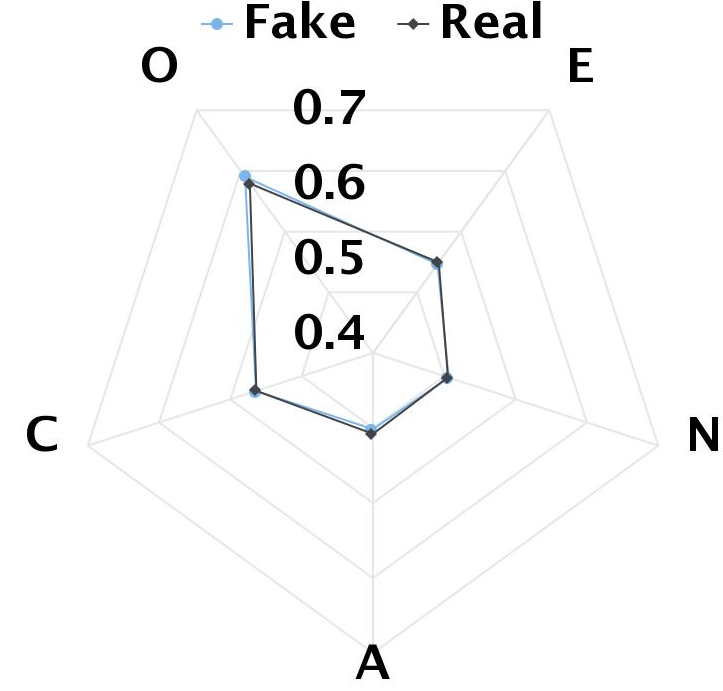}}
}
\caption{\textbf{Personality Comparison}. We measure personality with Five Factor Model and demonstrate the average values of these dimensions. }\label{fig:per}\vspace{-0.3cm}
\end{figure}
To predict personality, we apply an unsupervised personality prediction tool named Pear~\cite{celli2014pr2}, a state-of-the-art  text-based personality prediction model. Pear provides a pre-trained model using the user historical tweets\footnote{\url{http://personality.altervista.org/pear.php}}. The results are shown in Figure~\ref{fig:per}, where we can see that on both datasets: users in $\mathcal{U}^{(f)}$ tend to have relatively low Neuroticism, indicating less anxiety. Interestingly, this observation contradicts with the findings that emotion and anxiety makes people prone to fake news and disinformation~\cite{difonzo2007rumor}. It is probably because that we use tweets to characterize an individual's ``online personality'', which could be biased by the tweets and show different results with the his/her real personality.

\begin{figure}[tb!]
\centering
\subfigure[Location on PolitiFact]{
  {\includegraphics[scale=0.122]{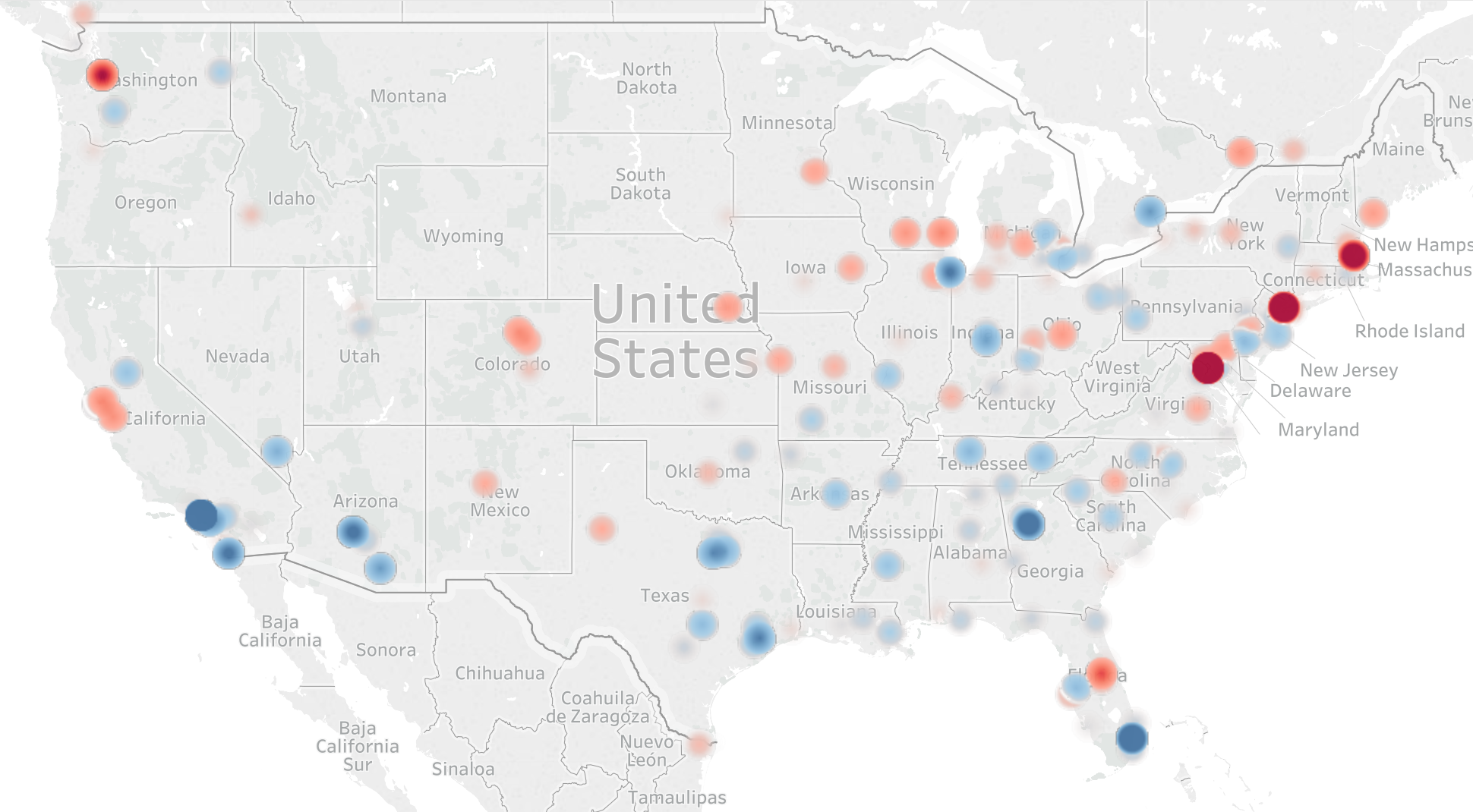}}
}
\subfigure[Location on Gossipcop]{
  {\includegraphics[scale=0.10]{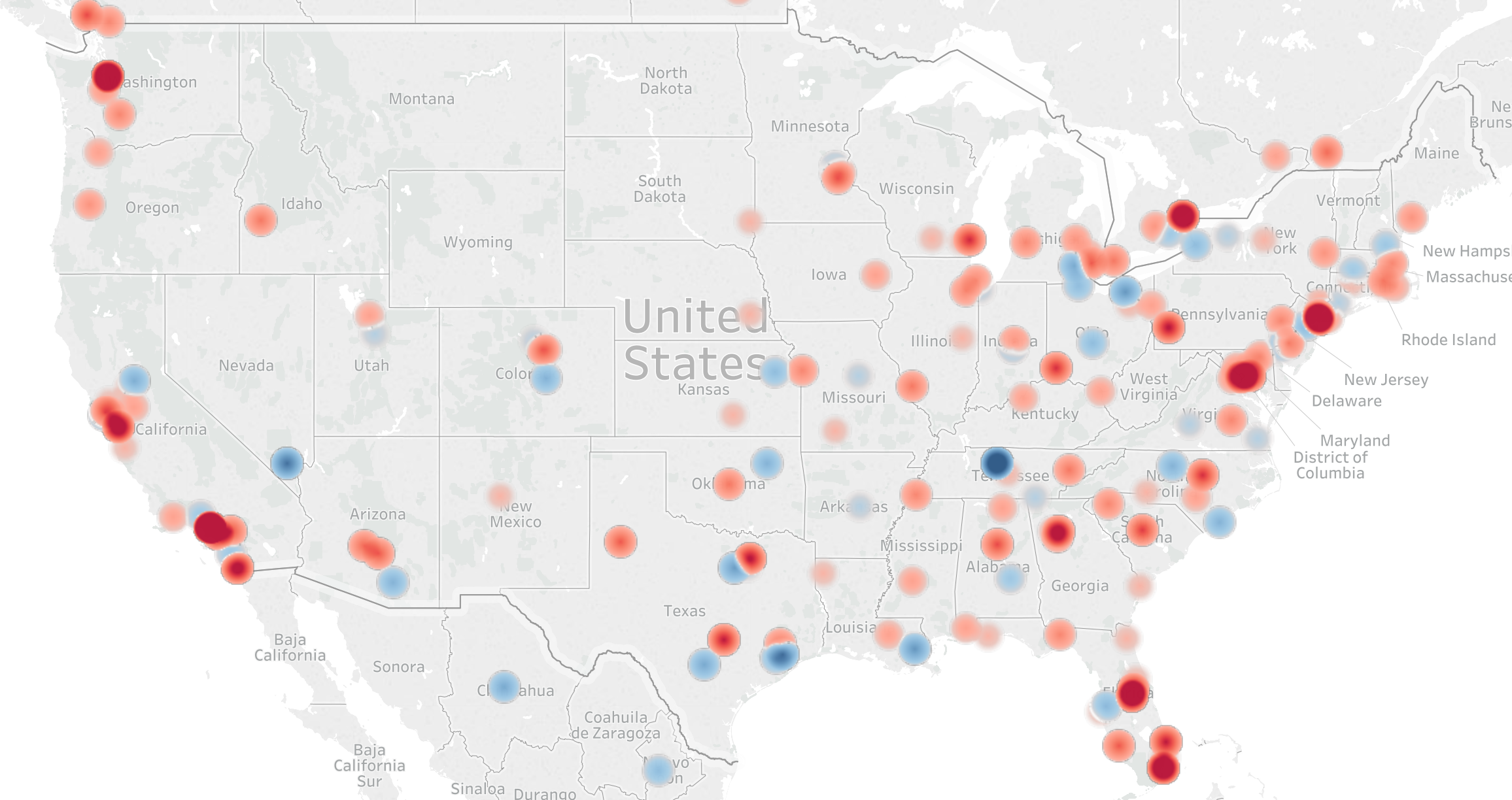}}
}
\caption{\textbf{Location Comparison}. Each point in the map denotes the difference of number of tweets posted by users. Red indicates more real news and blue shows more fake news.}\label{fig:loc}\vspace{-0.4cm}
\end{figure}

\textbf{Location:} Research has shown an inseparable relationship between user profiles and geo-locations. Even though users are allowed to reveal their geo-location explicitly, the location fields are very sparse, noisy, and unstructured~\cite{cheng2010you}. Thus, we exploit user-posted content to predict the user's location~\cite{bogeolocation,wu2014toward,rahimi2016pigeo}. The idea is to identify ``location indicative words'' (LIW), which can implicitly or explicitly encode an association with a particular location. The implementation of a pre-trained LIW model is integrated into an open source tool named pigeo~\cite{rahimi2016pigeo}, which is utilized here to predict the geo-locations of users in $\mathcal{U}^{(f)}$ and $\mathcal{U}^{(r)}$. The predicted results of pigeo are at the city-level and also include $(latitude,longitude)$ pairs, from Figure~\ref{fig:loc} we observe that: (1) there are overall more users located in the US than other places, which is because most of the real/fake news items in our particular datasets are published and related to US politics and entertainments; and (2) the location distribution is different for fake and real news on both datasets, and the red and blue dots demonstrate the degree of differences. For example, there are general more real news share in east region of US in our datasets.


\begin{figure}[!tb]
\centering
\subfigure[Profile Image on PolitiFact]{
  {\includegraphics[scale=0.26]{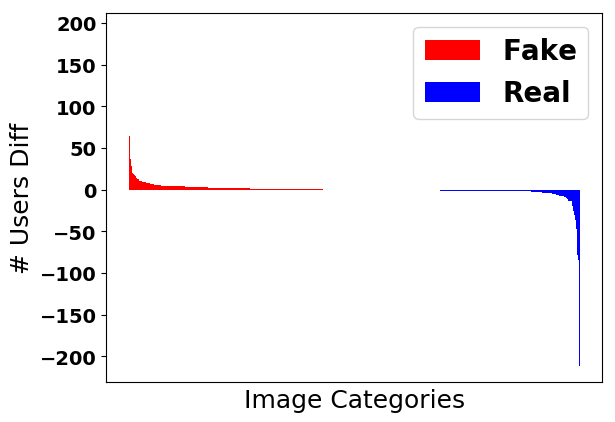}}
}
\hspace{-0.2cm}
\subfigure[Profile Image on GossipCop]{
  {\includegraphics[scale=0.25]{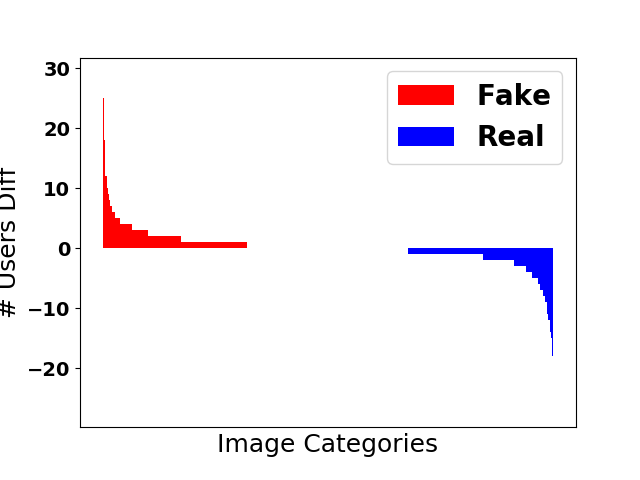}}
}
\caption{\textbf{Profile Image Comparison}. The x-axis indicates the 1000 image category labels from ImageNet. The y-axis represents the difference of the user counts for fake news to the user counts for real news.}\label{fig:img}

\end{figure}

\textbf{Profile Image:} Profile images are important visual components of users on social media. Various studies have demonstrated the correlation between the choice of profile images with user personalities~\cite{liu2016analyzing}, behaviors, and activities~\cite{tominaga2015study}. We aim to answer the following question: do users use different types of profile images among those who are more likely to share fake news versus real news? To test this hypothesis, we classify the object types in profile images. With the recent development of deep learning in the computer vision domain, convolutional neural networks (CNN) have shown good performance for detecting objects in images. We chose the pre-trained VGG16 model~\cite{sohn2015learning} as it is the widely-used CNN architecture. 
The resultant distribution of object types in profile images are shown in Figure~\ref{fig:img}.
We can see that: the distributions of profile image classes are different for users in $\mathcal{U}^{(f)}$ and $\mathcal{U}^{(r)}$ on both datasets. For example, there are specific image types\footnote{\url{http://image-net.org/explore}}, such as ``wig'' and ``mask'' dominating the image categories for users spreading fake news, and ``website'' and ``envelope'' dominating the image categories for users spreading real news, on both datasets consistently.

\textbf{Political Bias:} Political bias plays an important role in shaping users' profiles and affecting their news consumption choices on social media. Sociological studies on journalism demonstrate the correlation between  partisan bias and  news content authenticity (i.e., fake or real news)~\cite{gentzkow2014media}. Thus, we aim to answer the following questions: (1) are users with stronger political bias more likely to share fake news than real news; and  (2) are there clear differences in users' political bias distribution in fake and real news social engagements?

\begin{figure}[t!]
\centering
\subfigure[Political Bias on PolitiFact ]{
  {\includegraphics[scale=0.09]{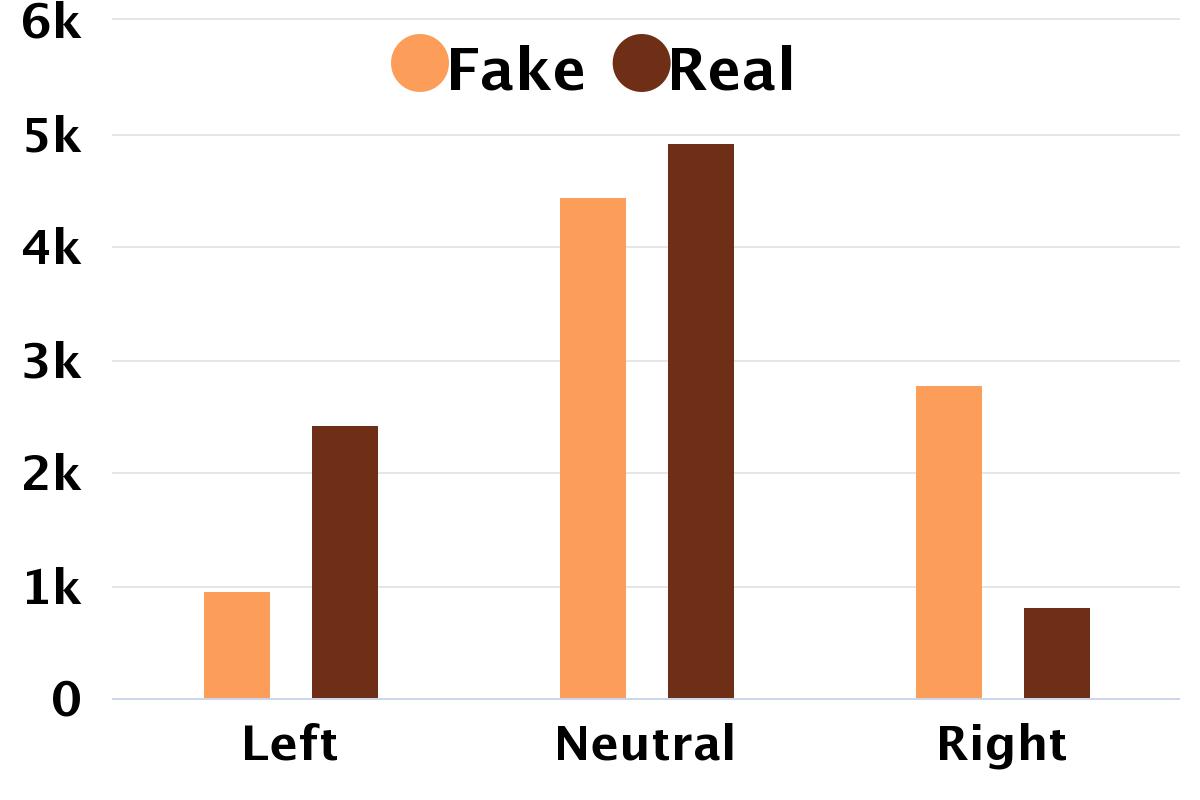}}
}
\subfigure[Political Bias on GossipCop]{
  {\includegraphics[scale=0.09]{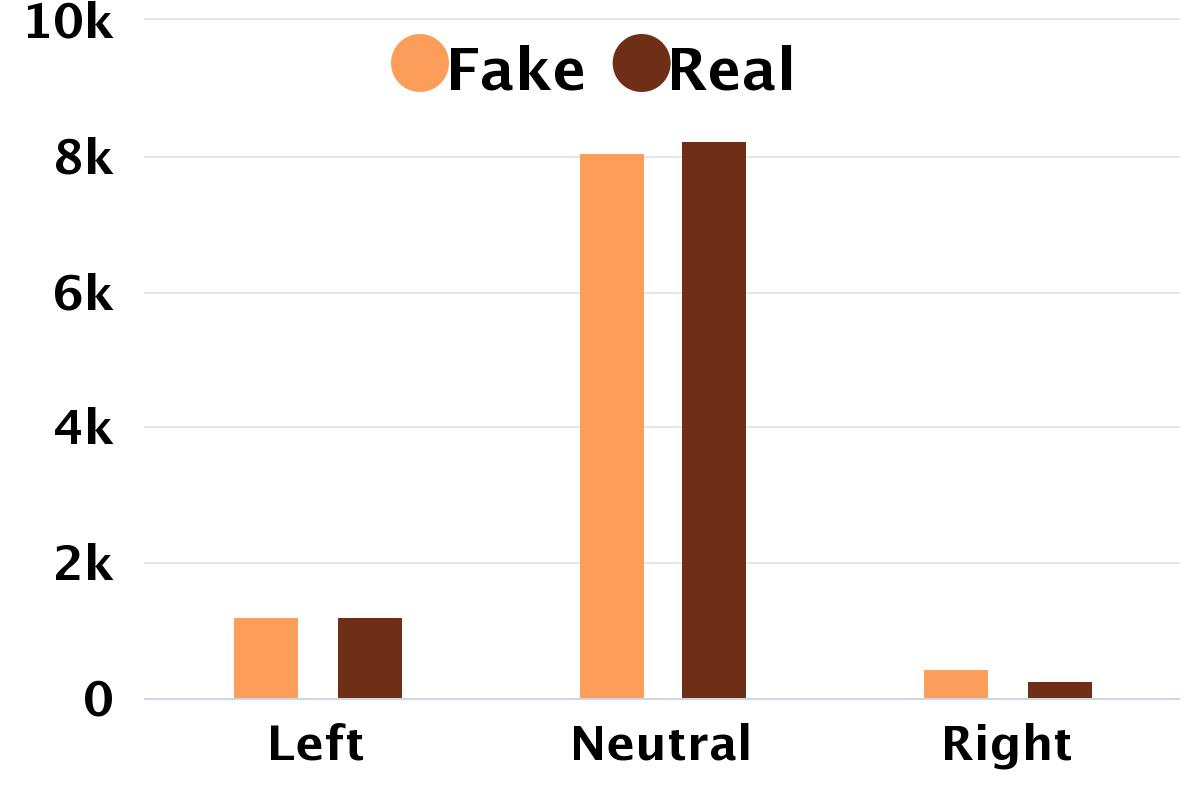}}
}
\caption{\textbf{Political Bias Comparison}. We plot the distribution of number of users with ``Left'', ``Neutral'', and ``Right'' political bias for both datasets.}\label{fig:bias}
\vspace{-0.3cm}
\end{figure}
To answer these questions, we need to measure the political bias scores of all users. Reports have shown people's political affiliation is correlated with their attributes and behaviors\footnote{\url{https://2012election.procon.org/view.resource.php?resourceID=004818}}. Thus, we adopt method in~\cite{kulshrestha2017quantifying} to measure user political bias scores by exploiting users' interests. The basic idea is that users who are more left-leaning or right-leaning have similar interests among each other. 
The resultant scores range from $[-1,1]$, where -1 indicates the most  right-leaning (republican-leaning),  1 represents most left-leaning (democrat-leaning), and 0 indicates the least-biased. We empirically set $0.5$ as our threshold such that scores that within $[-1,-0.5]$, $(-0.5,0.5)$, $(0.5,1]$ are treated as right-, neutral, and left-leaning. 
We observe that: (1) users that are more likely to share fake news (i.e., $u \in \mathcal{U}^{(f)}$)  also have a high probability to be biased on both datasets, and are more likely to be right-leaning; (2) users that are more likely to share real news (i.e., $u \in \mathcal{U}^{(r)}$) tend to be neutral-biased; and (3) overall,  users in the two datasets demonstrate different political bias score distributions, indicating that the  political bias of users could potentially help differentiate fake/real news.

\subsection{Explicit Profile Features}
A list of representative profile attributes include:
\begin{itemize}
\item Basic user description fields (\textbf{Profile-Related})\\
$\bullet$ \textit{Verified}:whether this is a verified user;\\
$\bullet$ \textit{RegisterTime}: the number of days  since the accounted was registered;
\item Attributes of user activities (\textbf{Content-Related})\\
$\bullet$ \textit{StatusCount}: the number of posts;\\
$\bullet$ \textit{FavorCount}: the number of favorites;
\item  Social networks attributes (\textbf{Network-Related})\\
$\bullet$ \textit{FollowerCount}: the number of followers;\\
$\bullet$ \textit{FollowingCount}: the number of users being followed.\\
\end{itemize}
We compare these six fields and test whether the users in $\mathcal{U}^{(r)}$ and $\mathcal{U}^{(f)}$ have clear differences. 
For categorical features such as \textit{Verified}, we demonstrate the ratio different categories; for numerical features such as \textit{RegisterTime}, we adopt the box-and-whisker diagram to show the key  quartiles of aggregated statistics. Due to the space limitation and better organization, we show the figures for \textit{RegisterTime}. Other features are analyzed similarly and omitted here, but are included here\footnote{The omitted figures are available at \url{https://tinyurl.com/y5mmdj2u} }.

\begin{figure}[tbp]
\centering
\subfigure[RegisterTime on PolitiFact]{
  {\includegraphics[scale=0.43]{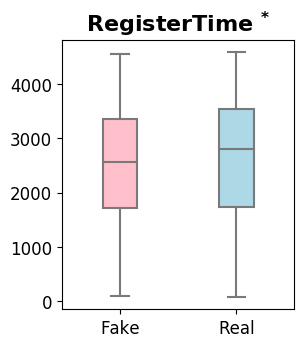}}
}
\subfigure[RegisterTime on GossipCop]{
  {\includegraphics[scale=0.44]{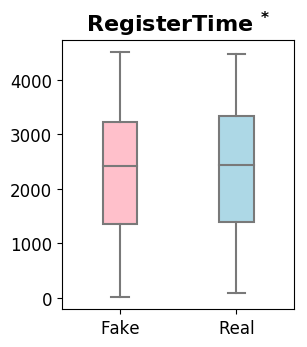}}
}
\caption{\textbf{Profile Features Comparison}. We show the Box-Plot to demonstrate the distribution of \textit{RegisterTime} for users.
}\label{fig:user_pro}
\vspace{-0.3cm}
\end{figure}

Profile-related features are compared in Figure~\ref{fig:user_pro}. 
We count the number of verfied and unverified users on both datasets, and observe that there are $938$ and $188$ more verified users in $\mathcal{U}^{(r)}$ than $\mathcal{U}^{(f)}$ on PolitiFact and GossipCop, which shows that verified users are more likely to share real news. 
We rank RegisterTime values of all users in $\mathcal{U}^{(f)}$ and $\mathcal{U}^{(r)}$ and perform a two-tail statistical $t$-test, and the $p$-value is less that 0.05. In addition, the box-and-whisker diagram shows that the distribution of user RegisterTime exhibits a significant difference between these user groups. The observations on both datasets demonstrate that users who are more likely to share fake news registered approximately $132$ and $49$  days earlier than those users in $\mathcal{U}^{(r)}$. This is consistent with previous work~\cite{castillo2011information}, which is because new accounts are intentionally created to spread fake news that are collected in recent years.

For content-related profile features, we rank the StatusCount and FavorCount and perform a t-test, yielding the same observations on both datasets: (1)  users in $\mathcal{U}^{(f)}$ generally publish fewer posts than users in $\mathcal{U}^{(r)}$, which indicates those users sharing more real news are likely to be more active; and (2) users in $\mathcal{U}^{(f)}$ tend to express more ``favor'' actions to tweets posted by other users, indicating their willingness to reach out to other users.



For network-related profile features, we rank the FollowerCount and FollowingCount and perform a t-test, and observe that user in $\mathcal{U}^{(f)}$ have fewer followers and more following counts from on both datasets significantly. For example, there are $46$ and $579$ fewer followers in $\mathcal{U}^{(f)}$ for PolitiFact and GossipCop, respectively.

In summary, we conclude that users in $\mathcal{U}^{(f)}$ and $\mathcal{U}^{(r)}$ reveal different feature distributions in most explicit and implicit feature fields, answering \textbf{RQ2}. These observations have great potential to guide the fake news detection process, which will be explored in detail in the next section.



\section{Exploiting User Profiles}\label{sec:eval}
In this section, we address \textbf{RQ3}. We explore whether the  user profile features can help improve fake news detection, and how we can build effective models based on them, with feature importance and model robustness analysis.
\begin{table*} [htbp!]
\centering \caption{Performance comparison for fake news detection with different feature representations.}
\begin{tabular}{|l|l|c|c|c|c|c|}
\hline
Datasets & Metric  &RST & LIWC & UPF & RST\_UPF & LIWC\_UPF  \\
\hline \hline
\multirow{4}{*}{\textbf{PolitiFact}}   & Accuracy  & 0.782 & 0.830 & 0.909 &0.918  &  0.921 \\
\cline{2-7}
&Precision &0.777 & 0.809 & 0.948 &0.949  &  0.942 \\
\cline{2-7}
&Recall&0.786 & 0.861 & 0.864 &0.883  &0.897  \\
\cline{2-7}
&F1 & 0.781 & 0.834 & 0.904 & 0.915 & 0.919 \\
\hline
\hline
\multirow{4}{*}{\textbf{GossipCop}} & Accuracy & 0.598 & 0.751 &0.966  &   0.966  &0.963  \\
\cline{2-7}
&Precision & 0.601 & 0.796 & 0.956 &  0.952 & 0.949 \\
\cline{2-7}
&Recall & 0.585 & 0.674 & 0.976 & 0.978  & 0.978 \\
\cline{2-7}
&F1 &  0.593&0.730  &  0.966&  0.967 & 0.963 \\
\hline
\end{tabular} \label{tab:performance}
\end{table*}

\subsection{Experimental Settings}

We first introduce how to extract user profile features $\mathbf{f}$ for news  $a$. Let $\mathcal{U}$ denote the set of users who share news $a$. For each user $u_i\in\mathcal{U}$, we extract all types of aforementioned profile features and concatenate them into one feature vector $\mathbf{u}_i$. Note that for profile image features, since it has 1000 types, we use Principle Component Analysis~\cite{jolliffe2011principal} to reduce the dimension to 10. Then we represent the user profile feature of a news as the average feature scores of all the users that share the news, i.e., $\mathbf{f}=\frac{1}{|\mathcal{U}|}\sum_{u_i\in\mathcal{U}}\mathbf{u}_i$. We also denote the proposed \textit{U}ser \textit{P}rofile \textit{F}eature vector $\mathbf{f}$ as UPF.


To evaluate the performance of fake news detection algorithms, we use the following commonly used metrics to evaluate classifiers: Accuracy, Precision, Recall, and F1. We randomly choose 80\% of news pieces for training and remaining 20\% for testing, and the process is performed for 5 times and the average performance is reported. We compare the proposed features UPF with several state-of-the-art  feature representations for fake news detection as below\footnote{All data and code are available at \url{https://tinyurl.com/y5mmdj2u} }:

\begin{itemize}
\item \textbf{RST}~\cite{ji2014representation}: RST can capture the writing style of a document by extracting the rhetorical relations systematically. It  learns a transformation from a bag-of-words surface representation into a latent feature representation\footnote{The code is available at: https://github.com/jiyfeng/DPLP}. 

\item \textbf{LIWC}~\cite{pennebaker2015development}: LIWC extracts lexicons that fall into different psycholinguistic categories, and learn a feature vector through multiple measures for each document\footnote{The software and description of measures are available at: http://liwc.wpengine.com/}.

\item \textbf{RST\_UPF}. RST\_UPF represents the concatenated features of RST and UPF, which includes features extracted from both news content and user profiles.
\item \textbf{LIWC\_UPF}. LIWC\_UPF represents the concatenated features of LIWC and UPF, which includes features extracted from both news content and user profiles.
\end{itemize}
We compare UPF with state-of-the-art fake news detection features that are extracted from \textbf{news content}, e.g., Rhetorical Structure Theory (RST) discourse parsing~\cite{ji2014representation} and Linguistic Inquiry and Word Count (LIWC)~\cite{pennebaker2015development}. We also combine RST and LIWC features with UPF to further explore if UPF have complementary information. 

\subsection{Fake News Detection Performance Comparison}

We test the baseline features on different learning algorithms, and choose the one that achieves the best performance (see Table~\ref{tab:performance}). The algorithms include Logistic Regression (LogReg for short), Na\"ive Bayes (NBayes), Decision Tree (DTree), Random Forest (RForest), and AdaBoost. We used the open-sourced \textit{scikit-learn}\footnote{https://scikit-learn.org/stable/} machine learning framework in Python to implement all these algorithms. To ensure a fair comparison of features, we ran all the algorithms using default parameter settings. We have the following observations:
\begin{itemize}
\item For news-content-based methods, we see that LIWC performs  better than RST. This indicates that the LIWC vocabulary can better capture the deceptiveness in news content, which reveals that fake news pieces are very different from real news in terms of word choice from psychometrics perspectives.

\item Our proposed UPF can achieve good performance in both datasets on all metrics. This shows that users that share more fake news and real news have  different demographics and characteristics on social media, which serve as good features for fake news detection.
\item In addition, RST\_UPF performs better than either RST or UPF, which reveals that they are extracted from orthogonal  information spaces, i.e., RST features are extracted from news content and UPF features from user profiles on social media, and have complementary information to help fake news detection.
\end{itemize}

\subsection{Choice of Learning Algorithms}\label{sec:classifier}
To evaluate the robustness of the extracted features UPF, we compare the fake/real news classification performances using different common classifiers, i.e., Random Forest (RF), Support Vector Machine (SVM), Decision Trees (DT), and Logistic Regression (LR).  The results are shown in Table~\ref{tab:learning}. These algorithms have different learning biases, and thus their performance is often  different for the same task. While we observe that: (1) RF achieves the best overall performance on both datasets; and (2) while the performance of RF is slightly better than other learning algorithms, the results are not significantly different across algorithms. This demonstrates that when sufficient information is available in the user profile features, the performance is not very sensitive to the choice of learning algorithms. 

\begin{table}[t!]
\centering
\caption{Detection Performance for UPF with Different Learning Algorithms}
\begin{tabular}{|l|c|c|c|c|c|}
\hline
  Dataset & Model & Acc & Prec & Recall & F1\\
\hline
\hline
\multirow{4}{*}{\textbf{PolitiFact}} & RF & \textbf{0.909}& \textbf{0.948}  & \textbf{0.864} & \textbf{0.904} \\
\cline{2-6}
&SVM & 0.869 & 0.884 & 0.847 &0.865 \\
\cline{2-6}
&DT&  0.848&  0.849 & 0.844 & 0.847\\
\cline{2-6}
&LR& 0.848& 0.867& 0.819&0.842 \\
\hline
\hline
\multirow{4}{*}{\textbf{GossipCop}} & RF & \textbf{0.966}  & \textbf{0.956} & \textbf{0.976} & \textbf{0.966}  \\
\cline{2-6}
&SVM & 0.918 & 0.920 &  0.916 & 0.918 \\
\cline{2-6}
&DT&  0.931 &0.931  & 0.930 & 0.931\\
\cline{2-6}
&LR & 0.914  &0.918  & 0.909  & 0.914 \\
\hline
\end{tabular} \label{tab:learning}
\vspace{-0.2cm}
\end{table}

\subsection{Feature Importance Analysis}
Now we analyze the relative importance of these features for predicting fake news. We analyze  feature importance in the Random Forest (RF)  by computing a feature importance score based on the Gini impurity~\cite{liaw2002classification}\footnote{\url{http://scikit-learn.org/stable/auto_examples/ensemble/plot_forest_importances.html}. A higher Gini impurity score indicates a higher importance}. The top 5 common important features (with Gini impurity scores) are:

\begin{enumerate}
\item \textit{RegisterTime} ($0.937$): the feature vector indicating the average distribution of verified and unverified users;
\item \textit{Verified} ($0.099$): the feature vector indicating the average distribution of verified and unverified users;
\item \textit{Political Bias} ($0.063$): the average bias score;
\item \textit{Personality} ($0.036$): the average distribution of users' personality scores characterized by five factors distribution;
\item \textit{StatusCount} ($0.035$): the average count of user posts.
\end{enumerate}

We observe that (1) RegisterTime is the most important feature because newly created account may be more likely for fake news propagation purpose; (2) the distribution of verified/unverified user counts is important as verified users are less likely to spread fake news (3) the average political bias score is important because those users who share fake news are more likely to be biased to a specific ideology, while users that share real news tend to be least biased;  (4) personality features are discriminative for detecting fake news because users' personalities affect their cognition and the way they respond to the real world~\cite{cantor2017personality}; and (5) the high importance score of StatusCount shows that the degrees of user activeness are quite different among users spreading fake and real news.

\begin{table}[tbp!]
\centering
\caption{Detection Performance with Different Group of Features from UPF}
\begin{tabular}{|l|c|c|c|c|c|}
\hline
  Dataset & Feature Group & Acc & Prec & Recall & F1\\
\hline
\hline
\multirow{4}{*}{\textbf{PolitiFact}} & All & \textbf{0.909}& \textbf{0.948}  & \textbf{0.864} & \textbf{0.904} \\
\cline{2-6}
&Explicit &  0.870  & 0.891 & 0.841 & 0.865\\
\cline{2-6}
&Implicit & 0.837 &0.892  & 0.763 & 0.823\\
\hline
\hline
\multirow{4}{*}{\textbf{GossipCop}} & All & \textbf{0.966}  & \textbf{0.956} & \textbf{0.976} & \textbf{0.966}  \\
\cline{2-6}
&Explicit&  0.894  & 0.884  &  0.906 & 0.895\\
\cline{2-6}
&Implicit  & 0.961  & 0.956 & 0.967  & 0.962\\
\hline
\end{tabular} \label{tab:feature_group}
\vspace{-0.2cm}
\end{table}

We further categorize the user profile features into three groups: \textbf{Explicit}, \textbf{Implicit} and \textbf{All} (i.e., both explicit and implicit features) and compare their contributions to the fake news detection task. The results are shown as in Table~\ref{tab:feature_group}. We observe that; (1) when all profile features are considered, the performance is higher than when only explicit or implicit features are considered. For example, the F1 scores on All features show a $4.51\%$  and $9.84\%$  increase compared with explicit and implicit feature groups on PolitiFact. This demonstrates that explicit and implicit features contain complementary information that can  improve detection performance. (2) The implicit feature group is much more effective than the explicit feature group on Gossipcop for Accuracy and F1 scores. Note that implicit features require user-generated content to infer their values, which requires more effort to construct, while explicit features are often directly available in users' raw data. These observations allow us to better balance the trade-off with limited time and resources to make more informed decisions when building these feature groups.

\section{Related Work}\label{sec:related}
In this section, we discuss the related work from two aspects: (1) fake news detection on social media; and (2) measuring user profiles on social media.

\subsection{Fake News Detection on Social Media}
Fake news detection approaches generally fall into two categories depending on whether they use (1) \textit{news content}; and (2)  \textit{social contexts}~\cite{shu2017fake}. For news content based approaches, features are extracted as linguistic-based such as writing styles~\cite{potthast2017stylometric}, and visual-based such as fake images~\cite{gupta2013faking}. 
Linguistic-based features capture specific writing styles and sensational headlines that commonly occur in fake news content~\cite{potthast2017stylometric}, such as lexical and syntactic features. Visual-based features try to identify fake images~\cite{gupta2013faking} that are intentionally created or capturing specific characteristics for images in fake news.
 News content based models include i) knowledge-based: using external sources to fact-checking claims in news content~\cite{magdy2010web,wu2014toward}, and 2) style-based: capturing the manipulators in writing style, such as deception~\cite{rubin2015truth} and non-objectivity~\cite{potthast2017stylometric}. 
Social context based approaches incorporate features from social media user profiles, post contents, and social networks. User features measure users' characteristics and credibility~\cite{castillo2011information}. Post features represent users' social responses, such as stances~\cite{jin2016news}. Network features are extracted by constructing specific social networks, such as diffusion networks~\cite{kwon2013prominent} or co-occurrence networks~\cite{ruchansky2017csi}. All of these social context models can basically be grouped as either stance-based or propagation-based. Stance-based models utilize users' opinions towards the news to infer news veracity~\cite{jin2016news}. Propagation-based models apply propagation methods to model unique patterns of information spread~\cite{jin2016news}.


Existing approaches that exploit user profiles simply extract features to train classifiers without a systematic understanding, which makes it a black-box that is difficult to interpret. Thus, to improve the explanatory power of fake new detection and to understand how to exploit user profiles to detect fake news, we perform, to our best knowledge, the first in-depth investigation of user profiles for their usefulness for fake news detection. 

\subsection{Measuring User Profiles on Social Media}
User profiles on social media generally contain both \textit{explicit} and \textit{implicit} features. Explicit profile features (e.g., post count), which are already provided in raw user meta data, are widely exploited in different tasks on social media, such as information credibility classification~\cite{castillo2011information} and user identity linkage~\cite{zafarani2013connecting}. While implicit profile features (e.g., personality), which are not directly provided, have proven very useful to apply to several specific analysis tasks. 
For age prediction, previous studies extract features from text posted by users~\cite{nguyen2013old}. Schwartz \textit{et al.} predicts gender, personality, and/or age simultaneously with open-vocabulary approaches~\cite{schwartz2013personality}. For political bias prediction, existing works rely on tweets and hashtags, network structure~\cite{conover2011predicting}, and language usage styles~\cite{makazhanov2014predicting}. Location prediction can be performed using ``Location Indicative Words'' (LIW)~\cite{rahimi2016pigeo}. Profile images can be predicted using a pre-trained model~\cite{deng2009imagenet} in an unsupervised manner.

We consider and extract both provided explicit and inferred implicit user profile features to better capture the different demographics of users for fake news detection.

\section{Conclusion and Future Work}\label{sec:conclude}
In this work, we aim to answer questions regarding nature and extent of the correlation between user profiles on social media and fake news and provide a solution to utilize user profiles to detect fake news.  Now, we summarize our findings of each research question and discuss future work.

\textbf{\textit{RQ1}} \textit{Which users are more likely to share fake news or real news?} To perform this study, we construct two real-world datasets, both including news content and social context, with reliable ground truth. We used both the absolute measure and relative measure to assess users' sharing behaviors in news, and identified  two sets of users representing those who are more likely to share fake/real news.

\textbf{\textit{RQ2}} \textit{What are the characteristics of users that are more likely to share fake/real news, and do they have clear differences?} With the two selected  sets of users, we explore their profiles from both explicit and implicit perspectives. We perform detailed statistical comparisons of these features and found that most of the features have distinct values/distributions. These findings pave the way to build related features to detect fake news.

\textbf{\textit{RQ3}} \textit{Can we use user profile features to detect fake news and how?} We endeavor to build learning algorithms to utilize them to detect fake news. We evaluate the effectiveness of the extracted user profile features by comparing them with several state-of-the-art baselines. We show that: (1) these features can make significant contributions to help detect fake news; (2) these features are overall robust to different learning algorithms and can consistently achieve good results; and (3)  in case of limited time or resources, one can implement a limited set of features and obtain reasonable good performance.

This work opens up the doors for many areas of research. First, we will investigate the potential and foundation of other types of user feature in a similar way, such as content features and social network features, for fake news detection. Second, we will further investigate the correlations between malicious accounts and fake news to perform jointly detecting malicious accounts and fake news pieces. Third, we will explore various user engagement behaviors such as reposts, likes, comments, to further understand their utilities for fake news detection.



{
\small
\bibliographystyle{IEEEtran}
\bibliography{latex8}

\begin{thebibliography}{10}
\providecommand{\url}[1]{#1}
\csname url@samestyle\endcsname
\providecommand{\newblock}{\relax}
\providecommand{\bibinfo}[2]{#2}
\providecommand{\BIBentrySTDinterwordspacing}{\spaceskip=0pt\relax}
\providecommand{\BIBentryALTinterwordstretchfactor}{4}
\providecommand{\BIBentryALTinterwordspacing}{\spaceskip=\fontdimen2\font plus
\BIBentryALTinterwordstretchfactor\fontdimen3\font minus
  \fontdimen4\font\relax}
\providecommand{\BIBforeignlanguage}[2]{{%
\expandafter\ifx\csname l@#1\endcsname\relax
\typeout{** WARNING: IEEEtran.bst: No hyphenation pattern has been}%
\typeout{** loaded for the language `#1'. Using the pattern for}%
\typeout{** the default language instead.}%
\else
\language=\csname l@#1\endcsname
\fi
#2}}
\providecommand{\BIBdecl}{\relax}
\BIBdecl

\bibitem{shu2017fake}
K.~Shu, A.~Sliva, S.~Wang, J.~Tang, and H.~Liu, ``Fake news detection on social
  media: A data mining perspective,'' \emph{KDD exploration newsletter}, 2017.

\bibitem{nyhan2010corrections}
B.~Nyhan and J.~Reifler, ``When corrections fail: The persistence of political
  misperceptions,'' \emph{Political Behavior}, vol.~32, no.~2, pp. 303--330,
  2010.

\bibitem{swift2016americans}
A.~Swift, ``Americans’ trust in mass media sinks to new low. gallup. com, 14
  september,'' 2016.

\bibitem{castillo2011information}
C.~Castillo, M.~Mendoza, and B.~Poblete, ``Information credibility on
  twitter,'' in \emph{Proceedings of the 20th international conference on World
  wide web}.\hskip 1em plus 0.5em minus 0.4em\relax ACM, 2011, pp. 675--684.

\bibitem{shu2018fakenewsnet}
K.~Shu, D.~Mahudeswaran, S.~Wang, D.~Lee, and H.~Liu, ``Fakenewsnet: A data
  repository with news content, social context and dynamic information for
  studying fake news on social media,'' \emph{arXiv preprint arXiv:1809.01286},
  2018.

\bibitem{davis2016botornot}
C.~A. Davis, O.~Varol, E.~Ferrara, A.~Flammini, and F.~Menczer, ``Botornot: A
  system to evaluate social bots,'' in \emph{WWW}, 2016, pp. 273--274.

\bibitem{ferrara2016rise}
E.~Ferrara, O.~Varol, C.~Davis, F.~Menczer, and A.~Flammini, ``The rise of
  social bots,'' \emph{Communications of the ACM}, vol.~59, no.~7, pp. 96--104,
  2016.

\bibitem{schwartz2013personality}
H.~A. Schwartz, J.~C. Eichstaedt, M.~L. Kern, L.~Dziurzynski, S.~M. Ramones,
  M.~Agrawal, A.~Shah, M.~Kosinski, D.~Stillwell, M.~E. Seligman \emph{et~al.},
  ``Personality, gender, and age in the language of social media: The
  open-vocabulary approach,'' \emph{PloS one}, vol.~8, no.~9, p. e73791, 2013.

\bibitem{mccrae1999age}
R.~R. McCrae, P.~T. Costa, M.~P. de~Lima, A.~Sim{\~o}es, F.~Ostendorf,
  A.~Angleitner, I.~Maru{\v{s}}i{\'c}, D.~Bratko, G.~V. Caprara,
  C.~Barbaranelli \emph{et~al.}, ``Age differences in personality across the
  adult life span: parallels in five cultures.'' \emph{Developmental
  psychology}, vol.~35, no.~2, p. 466, 1999.

\bibitem{sap2014developing}
M.~Sap, G.~Park, J.~Eichstaedt, M.~Kern, D.~Stillwell, M.~Kosinski, L.~Ungar,
  and H.~A. Schwartz, ``Developing age and gender predictive lexica over social
  media,'' in \emph{EMNLP'14}, 2014.

\bibitem{celli2014pr2}
F.~Celli and M.~Poesio, ``Pr2: A language independent unsupervised tool for
  personality recognition from text,'' \emph{arXiv preprint arXiv:1402.2796},
  2014.

\bibitem{difonzo2007rumor}
N.~DiFonzo and P.~Bordia, \emph{Rumor psychology: Social and organizational
  approaches}.\hskip 1em plus 0.5em minus 0.4em\relax American Psychological
  Association Washington, DC, 2007, vol.~1.

\bibitem{cheng2010you}
Z.~Cheng, J.~Caverlee, and K.~Lee, ``You are where you tweet: a content-based
  approach to geo-locating twitter users,'' in \emph{CIKM'10}.

\bibitem{bogeolocation}
H.~Bo and P.~C.~T. BALDWIN, ``Geolocation prediction in social media data by
  finding location indicative words.''

\bibitem{wu2014toward}
Y.~Wu, P.~K. Agarwal, C.~Li, J.~Yang, and C.~Yu, ``Toward computational
  fact-checking,'' \emph{Proceedings of the VLDB Endowment}, vol.~7, no.~7, pp.
  589--600, 2014.

\bibitem{rahimi2016pigeo}
A.~Rahimi, T.~Cohn, and T.~Baldwin, ``pigeo: A python geotagging tool,''
  \emph{Proceedings of ACL-2016 System Demonstrations}, pp. 127--132, 2016.

\bibitem{liu2016analyzing}
L.~Liu, D.~Preotiuc-Pietro, Z.~R. Samani, M.~E. Moghaddam, and L.~H. Ungar,
  ``Analyzing personality through social media profile picture choice.'' in
  \emph{ICWSM'16}.

\bibitem{tominaga2015study}
T.~Tominaga and Y.~Hijikata, ``Study on the relationship between profile images
  and user behaviors on twitter,'' in \emph{WWW'15}.

\bibitem{sohn2015learning}
K.~Sohn, H.~Lee, and X.~Yan, ``Learning structured output representation using
  deep conditional generative models,'' in \emph{Advances in neural information
  processing systems}, 2015, pp. 3483--3491.

\bibitem{gentzkow2014media}
M.~Gentzkow, J.~M. Shapiro, and D.~F. Stone, ``Media bias in the marketplace:
  Theory,'' National Bureau of Economic Research, Tech. Rep., 2014.

\bibitem{kulshrestha2017quantifying}
J.~Kulshrestha, M.~Eslami, J.~Messias, M.~B. Zafar, S.~Ghosh, K.~P. Gummadi,
  and K.~Karahalios, ``Quantifying search bias: Investigating sources of bias
  for political searches in social media,'' in \emph{CSCW'17}.

\bibitem{jolliffe2011principal}
I.~Jolliffe, \emph{Principal component analysis}.\hskip 1em plus 0.5em minus
  0.4em\relax Springer, 2011.

\bibitem{ji2014representation}
Y.~Ji and J.~Eisenstein, ``Representation learning for text-level discourse
  parsing,'' in \emph{ACL'2014}, vol.~1, 2014, pp. 13--24.

\bibitem{pennebaker2015development}
J.~W. Pennebaker, R.~L. Boyd, K.~Jordan, and K.~Blackburn, ``The development
  and psychometric properties of liwc2015,'' Tech. Rep., 2015.

\bibitem{liaw2002classification}
A.~Liaw, M.~Wiener \emph{et~al.}, ``Classification and regression by
  randomforest.''

\bibitem{cantor2017personality}
N.~Cantor and J.~F. Kihlstrom, \emph{Personality, cognition and social
  interaction}.\hskip 1em plus 0.5em minus 0.4em\relax Routledge, 2017, vol.~5.

\bibitem{potthast2017stylometric}
M.~Potthast, J.~Kiesel, K.~Reinartz, J.~Bevendorff, and B.~Stein, ``A
  stylometric inquiry into hyperpartisan and fake news,'' \emph{arXiv preprint
  arXiv:1702.05638}, 2017.

\bibitem{gupta2013faking}
A.~Gupta, H.~Lamba, P.~Kumaraguru, and A.~Joshi, ``Faking sandy: characterizing
  and identifying fake images on twitter during hurricane sandy,'' in
  \emph{WWW'13}.

\bibitem{magdy2010web}
A.~Magdy and N.~Wanas, ``Web-based statistical fact checking of textual
  documents,'' in \emph{Proceedings of the 2nd international workshop on Search
  and mining user-generated contents}.\hskip 1em plus 0.5em minus 0.4em\relax
  ACM, 2010, pp. 103--110.

\bibitem{rubin2015truth}
V.~L. Rubin and T.~Lukoianova, ``Truth and deception at the rhetorical
  structure level,'' \emph{Journal of the Association for Information Science
  and Technology}, vol.~66, no.~5, pp. 905--917, 2015.

\bibitem{jin2016news}
Z.~Jin, J.~Cao, Y.~Zhang, and J.~Luo, ``News verification by exploiting
  conflicting social viewpoints in microblogs.'' in \emph{AAAI}, 2016, pp.
  2972--2978.

\bibitem{kwon2013prominent}
S.~Kwon, M.~Cha, K.~Jung, W.~Chen, and Y.~Wang, ``Prominent features of rumor
  propagation in online social media,'' in \emph{ICDM'13}.\hskip 1em plus 0.5em
  minus 0.4em\relax IEEE, 2013, pp. 1103--1108.

\bibitem{ruchansky2017csi}
N.~Ruchansky, S.~Seo, and Y.~Liu, ``Csi: A hybrid deep model for fake news,''
  \emph{arXiv preprint arXiv:1703.06959}, 2017.

\bibitem{zafarani2013connecting}
R.~Zafarani and H.~Liu, ``Connecting users across social media sites: a
  behavioral-modeling approach,'' in \emph{KDD'13}.

\bibitem{nguyen2013old}
D.-P. Nguyen, R.~Gravel, R.~B. Trieschnigg, and T.~Meder, ``" how old do you
  think i am?" a study of language and age in twitter,'' 2013.

\bibitem{conover2011predicting}
M.~D. Conover, B.~Gon{\c{c}}alves, J.~Ratkiewicz, A.~Flammini, and F.~Menczer,
  ``Predicting the political alignment of twitter users,'' in
  \emph{SocialCom}.\hskip 1em plus 0.5em minus 0.4em\relax IEEE, 2011, pp.
  192--199.

\bibitem{makazhanov2014predicting}
A.~Makazhanov, D.~Rafiei, and M.~Waqar, ``Predicting political preference of
  twitter users,'' \emph{Social Network Analysis and Mining}, vol.~4, no.~1, p.
  193, 2014.

\bibitem{deng2009imagenet}
J.~Deng, W.~Dong, R.~Socher, L.-J. Li, K.~Li, and L.~Fei-Fei, ``Imagenet: A
  large-scale hierarchical image database,'' in \emph{CVPR'09}.

\end{thebibliography}
}

\end{document}